\documentclass[reprint,superscriptaddress,amsmath,amssymb,a4paperaps,pra,nofootinbib]{revtex4-1}
\usepackage{dcolumn,amssymb,amsmath,amsfonts,graphicx,latexsym,float,xcolor}
\usepackage{epstopdf}
\usepackage{nameref}
\usepackage{ulem}
\begin{document}

\title{Bunching-antibunching crossover in harmonically trapped few-body Bose-Fermi mixtures}
\author{J. Chen}
\email {jie.chen@physnet.uni-hamburg.de}
\affiliation{Zentrum f\"ur Optische Quantentechnologien, Universit\"at Hamburg, Luruper Chaussee 149, 22761 Hamburg, Germany}
\author{J. M. Schurer}
\email{jschurer@physnet.uni-hamburg.de}
\affiliation{Zentrum f\"ur Optische Quantentechnologien, Universit\"at Hamburg, Luruper Chaussee 149, 22761 Hamburg, Germany}
\affiliation{The Hamburg Centre for Ultrafast Imaging, Universit\"at Hamburg, Luruper Chaussee 149, 22761 Hamburg, Germany}
\author{P. Schmelcher}
\email{pschmelc@physnet.uni-hamburg.de}
\affiliation{Zentrum f\"ur Optische Quantentechnologien, Universit\"at Hamburg, Luruper Chaussee 149, 22761 Hamburg, Germany}
\affiliation{The Hamburg Centre for Ultrafast Imaging, Universit\"at Hamburg, Luruper Chaussee 149, 22761 Hamburg, Germany}
\date{\today}

\begin{abstract}
We investigate the ground state of a few-body Bose-Fermi mixture in a one-dimensional harmonic trap with varying interaction strengths and mass ratio. A bunching-antibunching crossover of the bosonic species for increasing interspecies' repulsion is observed within our fully correlated \textit{ab~initio} studies. Interestingly, this crossover is suppressed if the bosonic repulsion exceeds a critical value which strongly depends on the mass ratio. In order to unveil the physical origin of this crossover, we employ different levels of approximations: while a species mean-field approach can account for the antibunching, only the inclusion of the interspecies correlations can lead to the bunching. We show that these correlations effectively create an induced bosonic interaction, which in turn elucidates the occurrence of the bosonic bunching. Finally, we derive a two-site extended Bose-Hubbard model which reveals the low-energy physics of the bosons for the case of much heavier fermions.
\end{abstract}

\maketitle
\section{Introduction}
Experimental achievements in ultracold atomic systems opened up a new era for the studies of quantum many-body systems \cite{cold_atom_rev}. Owning to the extraordinary controllability of the trapping geometries as well as the atomic interaction strengths, experiments allow for investigations of enormous diversities of aspects of ultracold atomic ensembles. A prominent example is the observation of the superfluid to Mott insulator phase transition of bosons in optical lattices \cite{BH_exp_1,BH_exp_2,BH_exp_3}. 

With the aid of sympathetic cooling, the realization of ultracold atomic mixtures has been put forward \cite{mixture_exp_bb_1,mixture_exp_bb_2,mixture_exp_bb_3,mixture_exp_bb_4,mixture_exp_bb_5,mixture_exp_bf_1,mixture_exp_bf_2,mixture_exp_bf_3,mixture_exp_bf_4,mixture_exp_bf_5,mixture_exp_bf_6,mixture_exp_ff_1,mixture_exp_ff_2,mixture_exp_ff_3,mixture_exp_ff_4,mixture_exp_ff_5}. In comparison to a single-species system, the interplay between intra- and interspecies interactions can provide features which are inaccessible for a single species. Examples are the phase separation in a dual-species BEC \cite{mixture_exp_bb_1}, the collapse of the degenerate Fermi gas in a Bose-Fermi mixture \cite{mixture_exp_bf_5} and the celebrated BCS-BEC crossover in a Fermi-Fermi mixture \cite{mixture_exp_ff_4, mixture_exp_ff_5}. Those intriguing features also stimulated enormous efforts on the theoretical side, revealing, for example, the phase separation and phase diagram \cite{mixture_theo_bf_phase_1,mixture_theo_bf_phase_2,mixture_theo_bf_phase_3,mixture_theo_bf_phase_4,mixture_theo_bb_phase_1,mixture_theo_bb_phase_2,mixture_theo_ff_phase_1,mixture_theo_ff_phase_2}, the stability conditions \cite{mixture_theo_bf_stability_1,mixture_theo_bf_stability_2,mixture_theo_bf_stability_3,mixture_theo_bb_stability_1}, and the collective excitations \cite{mixture_theo_bf_collective_1,mixture_theo_bf_collective_2,mixture_theo_bf_collective_3} in various atomic mixtures. 

Among them, the one-dimensional (1D) Bose-Fermi mixtures are interesting in their own rights because, on one hand, the presence of two kinds of statistics among the particles can bring about significant differences on a ``macroscopic'' level as compared to the other types of mixtures, such as the density profiles and the stability condition \cite{BEC_Pethick,BEC_Stringari}. On the other hand, the 1D nature yields new features compared to the physics in higher dimensions. For instance, a 1D Bose-Fermi mixture can be described by the Gaudin-Yang model \cite{mixture_theo_GYmodel_1,mixture_theo_GYmodel_2} in the strongly interacting Tonks-Girardeau (TG) regime, with the exact eigenstates obtained by the Bose-Fermi mapping \cite{mixture_theo_bf_TG_1,mixture_theo_bf_TG_2}. Besides, the low-energy physics can be described by the Tomonaga-Luttinger liquid theory making it resemble a liquid of polarons \cite{mixture_theo_bf_LL_1,mixture_theo_bf_LL_2,mixture_theo_bf_LL_3}.

Stimulated by the recent experimental progresses on few-body ensembles \cite{few_exp1,few_exp2, few_exp3, few_exp4, few_exp5,few_exp6}, significant theoretical effort also focuses on the 1D few-body mixtures \cite{few_bf_density1,few_bf_density2,few_bf_density3,few_bf_density4,few_bb_density1,few_quench1,few_quench2,few_quench3,few_quench4,few_bf_SC1,few_bf_SC2,few_bf_SC3}, revealing for example the density profiles and correlation functions \cite{few_bf_density1,few_bf_density2,few_bf_density3,few_bf_density4,few_bb_density1}, dynamical properties \cite{few_quench1,few_quench2,few_quench3,few_quench4}, and the equivalence to a spin-chain model in the strongly interacting regime \cite{few_bf_SC1,few_bf_SC2,few_bf_SC3}. Note, however, that while most of the discussions focus on either the strongly interacting limit or are limited to certain observables, studies which systematically explore the mixture properties from the many-body perspective are still rare \cite{few_bb_density1,few_quench1,few_quench2}. 

In the present work, we investigate the few-body ensemble of a 1D ultracold Bose-Fermi mixture with harmonic confinement. Particular emphasis is put on how the interactions and the mass ratio affect the system. The discussions cover a large range of varying mass ratio rather than a specific situation in the strongly mass-imbalanced regimes \cite{few_quench3,few_quench4}. Moreover, we have a focus on an effective intraspecies description from the perspective of the interspecies correlations. To this end, we employ the recently developed \textit{ab~initio} multilayer multiconfiguration time-dependent Hartree method for mixtures (ML-MCTDHX) \cite{MX,MB_01,MB_02}. By means of the imaginary time propagation \cite{imagine_time}, it enables us to obtain the ground-state wave function which takes all correlations (both intra- and interspecies correlations) into account.  We first present our main observation which is the bunching-antibunching crossover of the bosonic species encoded, e.g., in the reduced two-body density. In particular, we find an increased bosonic bunching tendency with increasing interspecies Bose-Fermi repulsion. Interestingly, the bunching process is suppressed above a critical value of the repulsive bosonic interaction. In the latter regime the increase of interspecies interaction leads to a bosonic antibunching within the parameter window studied and this critical bosonic repulsion reveals a strong mass imbalance dependency. In order to elucidate the physical origin of this crossover, we apply two approximate methods yielding direct insights into the structure of the many-body wave function and the mechanism of the crossover. First, we adopt a species mean-field (SMF) description excluding all the interspecies correlations. Through the buildup of a mean-field induced potential, the SMF description can qualitatively account for the antibunching regime, while it fails to describe the bunching regime. Second, we employ a beyond SMF description developed in Ref. \cite{effective_theory}, which accounts for the interspecies correlations to first order. In this way, we arrive at an effective single-species Hamiltonian containing, besides the induced potential, an additional induced Bose-Bose interaction. Importantly, such an induced interaction successfully explains the occurrence of the bosonic bunching. Finally, in the framework of the beyond SMF description, we derive a two-site extended Bose-Hubbard model which directly reveals the low-energy physics present among the bosons in the case where the fermions are much heavier.

This paper is organized as follows. In Sec. \ref{Theoretical_framework}, we introduce our setup including the Hamiltonian and our computational approach. In Sec. \ref{Ground_state}, we first present our main observation: the bunching-antibunching crossover for the bosonic species. To elucidate the crossover mechanism, we adopt the SMF and the beyond SMF descriptions. Furthermore, based on the profile of the effective potential, we introduce a two-site extended Bose-Hubbard model which enables us to describe the low-energy physics effectively present among the bosons in the strong mass-imbalanced regimes. Finally, our conclusions and outlook are provided in Sec. \ref{Conclusions}.

\section{Theoretical framework } \label{Theoretical_framework} 
In this section, we briefly describe our setup including the Hamiltonian and our computational approach. We focus on a few-body ensemble consisting of two bosons and two fermions in a one-dimensional harmonic trap. The ground state of this mixture is obtained from \textit{ab~initio} ML-MCTDHX simulations which include all correlations.

\subsection{Hamiltonian}
The Hamiltonian of our 1D harmonically trapped ultracold Bose-Fermi mixture is given by $\hat{H} = \hat{H}_{b}+\hat{H}_{f}+\hat{H}_{bf}$, where
\begin{align}
\hat{H}_{b} &=\int dx_{b}~\hat{\psi}^{\dagger}_{b}(x_{b}) \textit{h}_{b}(x_{b}) \hat{\psi}_{b}(x_{b}) \nonumber\\
&+\frac{g_{b}}{2}\int dx_{b}~\hat{\psi}^{\dagger}_{b}(x_{b})\hat{\psi}^{\dagger}_{b}(x_{b})\hat{\psi}_{b}(x_{b})\hat{\psi}_{b}(x_{b}), \nonumber \\
\hat{H}_{f} &=\int dx_{f}~\hat{\psi}^{\dagger}_{f}(x_{f}) \textit{h}_{f}(x_{f}) \hat{\psi}_{f}(x_{f}), \nonumber\\
\hat{H}_{bf} &= {g_{bf}}\int dx~\hat{\psi}^{\dagger}_{f}(x) \hat{\psi}^{\dagger}_{b}(x) \hat{\psi}_{b}(x) \hat{\psi}_{f}(x), \label{Hamiltonian_bf}
\end{align}
and $\textit{h}_{\sigma}(x_{\sigma}) = -\frac{\hbar^{2}}{2 m_{\sigma}}\frac{\partial^{2}}{\partial x_{\sigma}^{2}}+\frac{1}{2} m_{\sigma} \omega^{2} x _{\sigma}^{2}$ is the single-particle Hamiltonian for the harmonic confinement of the $\sigma = b(f)$ species. 
$ \hat{\psi}_{\sigma}^{\dagger}(x_{\sigma})$ [$\hat{\psi}_{\sigma}(x_{\sigma})$] is the field operator that creates (annihilates) a $\sigma$-species particle at position $x_{\sigma}$. For simplicity, we consider here equal trapping frequencies for both species. Moreover, we assume that the Bose-Bose (intraspecies) as well as the Bose-Fermi (interspecies) interactions are of zero range and can be modeled by contact potentials of strengths \cite{Feshbach_1,few_quench3}
\begin{align}
g_{b} = \frac{4 \hbar^{2}a_{b}}{m_{b} a_{\bot, b}^{2}}[1-C \frac{a_{b}}{a_{\bot, b}}]^{-1}, \\
g_{bf} = \frac{2 \hbar^{2}a_{bf}}{\mu a_{\bot, bf}^{2}}[1-C \frac{a_{bf}}{a_{\bot, bf}}]^{-1}.
\end{align}
Here $a_{b}$ ($a_{bf}$) is the 3D Bose-Bose (Bose-Fermi) $s$-wave scattering length and $C \approx 1.4603$ is a constant. The parameters $a_{\bot, b} = \sqrt{2 \hbar/ m_{b} \omega_{\bot}}$ and $a_{\bot, bf} = \sqrt{ \hbar/ \mu \omega_{\bot}}$ describe the transverse confinement, with $\mu = m_{b}m_{f}/({m_{b}+m_{f}})$ being the reduced mass and we assume the transverse trapping frequency $\omega_{\bot}$ to be equal for both species. Moreover, we focus on the repulsive interaction regime, i.e., $g_{b}$ $(g_{bf}) \geqslant 0$. It should be pointed out that, due to the Pauli-exclusion principle, the $s$-wave contributions to the fermionic scattering vanish and, hence, (spin-polarized) fermions become noninteracting at low collision energies. In the following discussion, we rescale the Hamiltonian \eqref{Hamiltonian_bf} for the units of the energy and length with $\eta = \hbar \omega$ and $\xi = \sqrt{\hbar /m_{b} \omega}$, respectively. We explore a mixture made of two fermions and two bosons, i.e., $N_{f} = N_{b} = 2$ and investigate the ground-state properties in both the mass-balanced ($\beta = 1$) and the mass-imbalanced ($\beta = 5,25$) regimes, with $\beta = m_{f}/m_{b}$ being the mass ratio. Let us note that such a 1D mixture is experimentally accessible by imposing strong transverse and weak longitudinal confinement for a binary mixture made of ${}^{7}\textrm{Li}$-${}^{6}\textrm{Li}$ ($\beta \approx 1$) \cite{mixture_exp_bf_1,mixture_exp_bf_2}, ${}^{171}\textrm{Yb}$-${}^{39}\textrm{K}$, ${}^{40}\textrm{K}$-${}^{7}\textrm{Li}$ ($\beta \approx 5$) or ${}^{171}\textrm{Yb}$-${}^{7}\textrm{Li}$ ($\beta \approx 25$). Moreover, the contact interaction strengths can be controlled experimentally by tuning the $s$-wave scattering lengths via Feshbach or confinement-induced resonances \cite{Feshbach_1,Feshbach_2,Feshbach_3}.

\subsection{Computational approach}
In order to explore our few-body system described by $\hat {H}$ from first principles, we employ the very recently developed ML-MCTDHX method \cite{MX, MB_01, MB_02}. Its efficient wave function representation scheme allows one to compute eigenstates as well as the temporal evolution of a many-body system including all correlations. 

To this end, the state of the 1D Bose-Fermi mixture $|\Psi (t) \rangle$ is first expanded as $|\Psi (t) \rangle = \sum_{i,j = 1}^{M} A_{ij}(t) |\psi_{i}^{f}(t) \rangle |\psi_{j}^{b}(t) \rangle$, with $\{|\psi_{i}^{\sigma}(t) \rangle\}_{i=1}^{i=M} $ being the states for $\sigma$ species, which form a set of orthonormal functions. It is important to note that both the coefficients $A_{ij}(t)$ and the species states $\{|\psi_{i}^{\sigma}(t) \rangle\}$ are time dependent. Furthermore, each species state $|\psi_{i}^{\sigma}(t) \rangle$ is expressed as linear combinations of the number states according to $|\psi_{i}^{\sigma}(t) \rangle = \sum_{\textbf{n}|N_{\sigma}} C_{i,\textbf{n}}^{\sigma}(t) |\textbf{n}\rangle^{\sigma}_{t}$, where $|\textbf{n} \rangle_{t}^{\sigma} = | n_{\sigma1}, n_{\sigma2}, \cdots \rangle$ are the number states for the $\sigma$ species under the constraint of particle number conservation $\sum_{i} n_{\sigma i} = N_{\sigma}$. Moreover, these number states $|\textbf{n} \rangle_{t}^{\sigma} $ are built by \textit{time-dependent} single-particle functions (SPFs) $\{| \phi_{k}^{\sigma}(t) \rangle\}_{k=1}^{s_{\sigma}}$. Using the Lagrangian variational principle results in the coupled integro-differential equations of motion for both coefficients $A_{ij}(t)$ and $C_{i,\textbf{n}}^{\sigma}(t)$ as well as the SPFs $|\phi_{k}^{\sigma}(t) \rangle$, which allows us to obtain the variationally optimized SPFs $| \phi_{k}^{\sigma}(t) \rangle$ and accordingly the state $|\Psi (t) \rangle$. 

Finally, we note that the numbers $s_{\sigma}$ and $M$ are the main control parameters for the numerical simulations, in which $s_{\sigma}$ truncates the dimension of the single-particle Hilbert space, leading to individual species spaces of size $K_{\sigma} = \left(\begin{array}{c}N_{\sigma}+s_{\sigma}-1 \\s_{\sigma}-1\end{array}\right) \left[K_{\sigma} = \left(\begin{array}{c}s_{\sigma} \\N_{\sigma}\end{array}\right) \right]$ for $N_{\sigma}$ bosons (fermions). The value of $M \leqslant \text{min} \{K_{A}, K_{B}\}$ defines how many species states are used to construct the full many-body Hilbert space.

\section{Ground-state properties and their effective descriptions} \label{Ground_state}
In this section, we investigate the ground-state properties of the mixture via the numerically obtained ground-state wave function of corresponding ML-MCTDHX simulations. The latter includes in principle all correlations. 
We begin with the presentation of our main result, the bunching-antibunching crossover of the bosonic species induced by the interaction with the fermions. This crossover is in particular signified in the reduced two-body density [see Eq.~\eqref{rho2_b}]. Having detected the crossover from the \textit{ab~initio} ML-MCTDHX simulations, we explore its physical origin by some effective descriptions. To do so, we first adopt a SMF description, which assumes the wave function of the mixture to be of product form taking into account the intraspecies correlations but excluding the interspecies correlations. While the SMF description can qualitatively describe the antibunching regime through the buildup of a mean-field induced potential, it fails to describe the bunching regime. Secondly, we go beyond the SMF approximation by including the interspecies correlations to first order \cite{effective_theory}. In this way, we arrive at an effective single-species Hamiltonian which contains, besides an induced potential, an additional induced interaction. Importantly, such an induced interaction unravels the profound physical insights ignored by the SMF description. In particular, it successfully explains the occurrence of the bosonic bunching. Finally, by adopting a single-band approximation in the strongly mass-imbalanced regime, we demonstrate the low-energy physics for the bosonic species via a two-site extended Bose-Hubbard model.

\subsection {Bosonic bunching-antibunching crossover}
\begin{figure}
  \centering
  \includegraphics[width=0.5\textwidth]{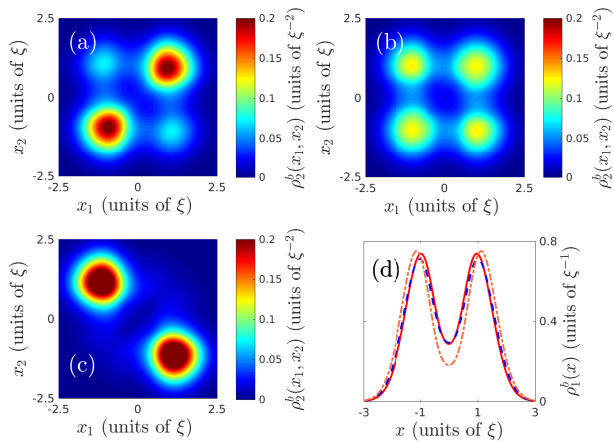}\hfill
  \caption{Spatial profiles of the bosonic reduced two-body density $\rho_{2}^{b} (x_1, x_2)$ for various $g_{b}$ with fixed $g_{bf} = 2.0$ and $\beta = 5$, in which (a) $g_{b} = 0$, (b) $g_{b} = 0.6$, and (c) $g_{b} = 2.0$. In addition, the corresponding profiles of the one-body density for the bosonic species are depicted in (d), where the red solid, blue dashed, and orange dash-dot line stands for $g_{b} = 0.0, 0.6, 2.0$, respectively.}
\label{p2_gb}
\end{figure}

\begin{figure}
  \centering
  \includegraphics[width=0.5\textwidth]{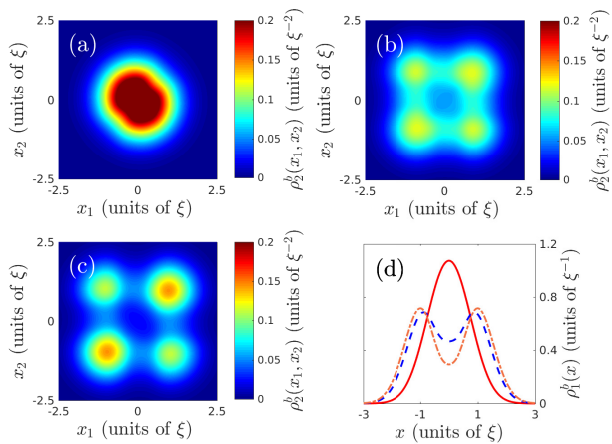}\hfill
  \caption{Spatial profiles of the bosonic reduced two-body density $\rho_{2}^{b} (x_1, x_2)$ for various $g_{bf}$ with fixed $g_{b} = 0.4$ and $\beta = 5$, in which (a) $g_{bf} = 0$, (b) $g_{bf} = 1.4$, and (c) $g_{bf} = 2.0$. In addition, the corresponding profiles of the one-body density for the bosonic species are depicted in (d), where the red solid, blue dashed, and orange dash-dot line stands for $g_{bf} = 0.0, 1.4, 2.0$, respectively.}
\label{p2_gbf}
\end{figure}

A particular feature which we observed for our mixture is the bunching-antibunching crossover for the bosonic species, which occurs as a function of the inter- and the intraspecies interaction strengths. Indeed, for a fixed $g_{bf}$ one detects a bunching to antibunching transition by increasing $g_{b}$ (cf. Fig.~\ref{p2_gb}), while for a fixed $g_{b}$ one finds the reverse transition for increasing $g_{bf}$ (cf. Fig.~\ref{p2_gbf}). In order to visualize this crossover, we introduce the reduced two-body density for the bosonic species 
\begin{equation}
\rho^{b}_{2} (x_1, x_{2}) = \langle \Psi | \hat{\psi}^{\dagger}_{b}(x_{1}) \hat{\psi}^{\dagger}_{b}(x_{2}) \hat{\psi} _{b}(x_{2}) \hat{\psi}_{b} (x_{1}) |\Psi \rangle. \label{rho2_b}
\end{equation}
The physical meaning of $\rho^{b}_{2} (x_1, x_{2})$ is the probability of finding one boson at position $x_1$ while the second one is at $x_2$, which naturally describes the spatial correlations between two bosons. Experimentally the spatial profile of the reduced two-body density can be measured via \textit{in~situ} absorption imaging (see \cite{in_situ} and references therein).

As an exemplary case, we elaborate how the bosonic species undergoes a transition from bunching to antibunching for fixed $g_{bf} = 2.0$ and increasing $g_{b}$. The reduced two-body densities are depicted in Fig.~\ref{p2_gb} for the mass-imbalanced case of $\beta=5$, together with the corresponding reduced one-body densities $\rho^{b}_{1} (x) = \int \rho_2^b(x,x') dx'$. Let us note that during the transition the two fermions are localized at the trap center due to the large mass ratio (see below). For the case $g_{b} = 0$, we observe that the bosons are bunching at either the left or right side of the trap, represented by the two dominant peaks around $x_1 = x_2 \approx \pm 1$ in the reduced two-body density. With increasing $g_{b}$, the bunching becomes energetically unfavorable, which immediately ramps down (up) the density in the vicinity of the diagonal ($x_{1} = x_{2}$) [off-diagonal ($x_{1} = -x_{2}$)] regions [cf. Fig.~\ref{p2_gb}(b)]. Finally, for the case $g_{b} = 2.0$, the bosons completely anti-bunch such that each boson resides on one side of the trap. Let us highlight that the reduced one-body density fails to capture the above crossover. For increasing $g_{b}$, we observe only a slight change of $\rho^{b}_{1} (x)$ with a minor broadening of the density distribution owing to the increment of the bosonic repulsion [cf. Fig.~\ref{p2_gb}(d)]. Since $\rho^{b}_{1} (x)$ is obtained by the partial trace of one particle over the corresponding reduced two-body density, which inevitably loses the two-particle correlations. Complementarily, the transition of the antibunching to the bunching behavior with the increase of $g_{bf}$ is presented as well (cf. Fig.~\ref{p2_gbf}). For increasing $g_{bf}$ and fixed $g_{b}$, we observe that the reduced two-body density evolves from the original profile with dominant populations on the off diagonal to the case of dominant populations on the diagonal. Similarly, this transition is not captured by the corresponding reduced one-body density as well, resulting in a deep dent of the density distribution from the original Gaussian profile with increasing the Bose-Fermi repulsion [cf. Fig.~\ref{p2_gbf}(d)]. 

In order to quantify the degree of the bosonic bunching (antibunching), we introduce the \textit{two-body density imbalance} as $\textit{P} = P_{s} -P_{d}$, where
\begin{align}
 P_{s}  &= 2 \int_{0}^{\infty} dx_{1} \int_{0}^{\infty} dx_{2} ~\rho^{b}_{2} (x_1, x_{2}) \nonumber \\
 P_{d}  &= 2 \int_{-\infty}^{0} dx_{1} \int_{0}^{\infty} dx_{2} ~\rho^{b}_{2} (x_1, x_{2}),
\end{align}
with $P_{s} (P_{d})$ being the probability of finding two bosons at the same side (different sides) of the trap. Thereby the two-body density imbalance directly reveals the probability difference between these two situations. Here, the prefactor $2$ originates from the parity symmetry of the Hamiltonian $\hat{H}$. Importantly, based on the observations for the spatial profiles of the bosonic reduced two-body densities, we arrive at the criteria of bosonic bunching (antibunching) as $\textit{P} >0$ ($\textit{P} <0$). Moreover, due to the normalization condition: $\int dx_{1} dx_{2} ~\rho^{b}_{2} (x_1, x_{2})  = 1$, the two-body density imbalance takes values within the interval $[-1,1]$. 

\begin{figure*}
  \centering
  \includegraphics[width=\textwidth]{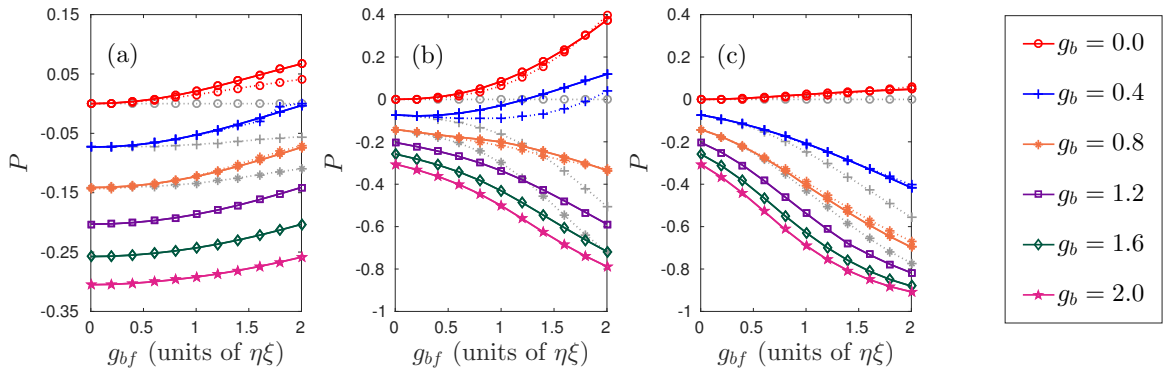}\hfill
  \caption{Two-body density imbalance for various interaction strengths and mass ratios, (a) $\beta = 1$, (b) $\beta = 5$, and (c) $\beta = 25$. All colored solid lines are the results obtained from the \textit{ab~initio} ML-MCTDHX simulations. In addition, the $P$ values for $g_{b} = 0.0, 0.4, 0.8$, obtained from the simulations of the SMF (gray dotted lines) and beyond SMF effective theory (colored dotted lines) are presented as well.}
\label{P_diff}
\end{figure*}

In Figs.~\ref{P_diff}(a)--\ref{P_diff}(c), we present the two-body density imbalance $P$ as a function of $g_{bf}$ for a set of discrete values of $g_{b}$ and various fixed values of $\beta$ (colored solid lines). Note, here, that the presented results are obtained from the  \textit{ab~initio} ML-MCTDHX simulations, which differ from the results using the adopted approximations and effective descriptions (see below). As expected, the increase of $g_{b}$ always reduces the $P$ value since the bosonic repulsion favors antibunching. However, the situation becomes more complicated once additionally $g_{bf}$ and $\beta$ are varied. For the equal-mass case, we observe that the increase of $g_{bf}$ leads to a monotonous increase of the $P$ value, whereas, in the mass-imbalanced regimes, an increase or decrease of the $P$ value depending on the value of $g_{b}$ being below or above a critical value $g_{b}^{c}$ takes place [cf. Figs.~\ref{P_diff}(b) and \ref{P_diff}(c)]. Interestingly, the $g_{b}^{c}$ decreases for increasing mass ratio, from $g_{b}^{c} \approx 0.6$ for $\beta = 5$ to $g_{b}^{c} \approx 0.2$ for $\beta = 25$. Besides, the mass ratio has also a significant impact on the absolute range of $P$ values with $P \in (-0.3,0.1)$ for $\beta =1$, while, becoming $P \in (-0.8,0.4)$ and $P \in (-0.9,0.1)$ for $\beta = 5$ and $25$, respectively.

\begin{figure}
  \centering
  \includegraphics[width=0.5\textwidth]{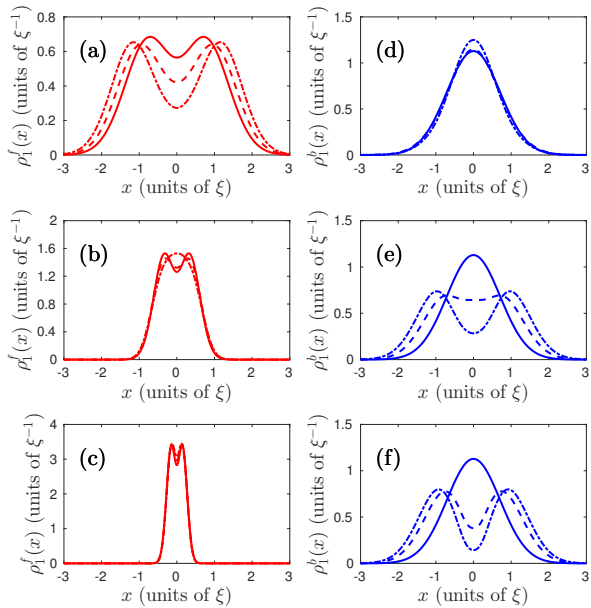}\hfill
  \caption{Spatial profiles of the one-body densities with fixed $g_{b} = 0$ for fermionic (a)--(c) and bosonic species (d)--(f). The upper, middle and lower panels are for the mass ratio $\beta = 1$, $\beta = 5$, and $\beta = 25$, respectively. Moreover, the solid, dashed, and dash-dot lines stand for the cases with $g_{bf} = 0$, $g_{bf} = 1.0$, and $g_{bf} = 2.0$, respectively. }
\label{p1_gbf}
\end{figure}

Before closing this section, let us briefly discuss the spatial profiles of the reduced one-body density $\rho^{\sigma}_{1} (x)$, which denotes the probability of finding a $\sigma$-species particle at position $x$.  In Fig.~\ref{p1_gbf}, we present $\rho^{\sigma}_{1} (x)$ for both species for changing $g_{bf}$ and $\beta$ and $g_{b} = 0$. Despite the fact that $\rho^{\sigma}_{1} (x)$ depends on the value of $g_{b}$ as well, we observe the increase of $g_{b}$ ($ g_{b} \in [0,2]$) only slightly affects these density profiles, resulting in a minor broadening of the bosonic density distribution owing to the increment of the bosonic repulsion (results are not shown here). For increasing $\beta$, we find the fermionic density distribution shrinks dramatically while the bosonic one is less affected. This observation can be quantitatively understood via the harmonic confinement encoded in the length scale $l_{\sigma} = \sqrt{\hbar /m_{\sigma} \omega}$ \cite{few_quench3}. A mass difference leads to different confinement lengths with the relative ratio $l_{b}/l_{f} = \sqrt{\beta}$, and thereby results in a smaller spatial overlap of the density profiles. By contrast, the increase of $g_{bf}$ has a large impact on both species, resulting in stronger demixing or phase separation of the two species. 

\subsection{Effective theoretical descriptions}
In order to elucidate the origin of the above-analyzed  bunching-antibunching crossover, hereafter, we present effective theoretical descriptions based on the fact that mixture is weakly entangled (see below). The ground-state wave function of the 1D Bose-Fermi mixture can be written in the form 
\begin{equation}
|\Psi \rangle = \sum_{i=1}^{\infty} \sqrt{\lambda_{i}} ~|\psi_{i}^{f}\rangle  |\psi_{i}^{b}\rangle, \label{psi}
\end{equation}
according to the Schmidt decomposition \cite{Schmidt}, where $ \lambda_{i} $ are the Schmidt numbers with descending order, i.e., \mbox{$\lambda_1 > \lambda_2 > \cdots$}, which are real positive numbers and obey the constraint $\sum_{i} \lambda_i = 1$ due to the normalization of the wave function $|\Psi \rangle$. Here $|\psi_{i}^{\sigma}\rangle $ denotes the $i$th Schmidt state for species $\sigma$. In addition, all the Schmidt states $\{|\psi_{i}^{\sigma}\rangle\} $ form an orthonormal basis. We emphasize that the Schmidt numbers directly reveal the interspecies correlations as $S = -\sum_{i} \lambda_{i} \text{log}_{2} \lambda_{i}$, with $S$ being the entanglement entropy \cite{Schmidt}. Moreover, the case $\lambda_{1} = 1$, i.e., $\lambda_{i \neq 1} = 0$, results in $S=0$, indicating that the mixture is nonentangled. 

\begin{figure*}
  \centering
  \includegraphics[width=\textwidth]{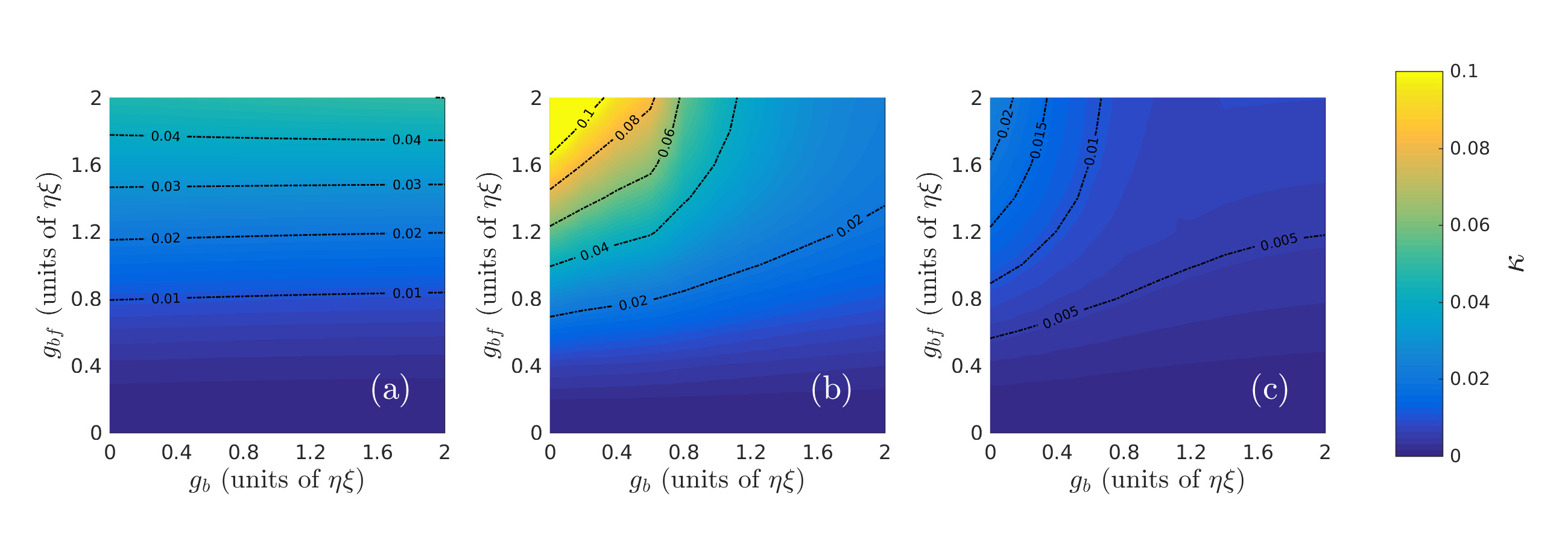}\hfill
  \caption{Species depletion $\kappa$ as a function of $g_{b}$ and $g_{bf}$ for (a) $\beta = 1$, (b) $\beta = 5$, and (c) $\beta = 25$, respectively.}
\label{species_depletions}
\end{figure*}

In order to quantitatively evaluate the interspecies correlations, we introduce the species depletion as $\kappa = 1-\lambda_{1}$, and depict it as a function of both $g_{b}$ and $g_{bf}$ for various fixed mass ratios (cf. Fig.~\ref{species_depletions}). Albeit the fact that the increase of $g_{bf}$ always ramps up the species depletion, however, the $\kappa$ value is also affected by both $g_{b}$ and $\beta$. For instance, for a fixed $g_{bf}$, the increase of $g_{b}$ significantly decreases the $\kappa$ value for $\beta = 5,25$, while it has minor impact on the case where the atoms possess the same mass, leaving the $\kappa$ value more or less unchanged. On the other hand, for fixed interaction strengths, the mass ratio can lead to dramatic variations of the species depletion. For the cases $g_{b} = 0$ and $g_{bf} = 2.0$, the $\kappa$ value is $0.134$ ($\beta=5$), $0.047$ ($\beta = 1$), and $0.024$ ($\beta=25$). Finally, we stress that the investigated Bose-Fermi mixture essentially remains within the weak entanglement regime ($\kappa \ll1$); here about $100\%$, $86\%$, and $100\%$ of the parameter space for $\beta = 1,5,$ and $25$, respectively,  exhibit a species depletion below $\kappa = 0.05$. It is this fact that we can exploit in the following to obtain an effective theoretical description for the bosonic species accounting for the bunching-antibunching crossover.

\subsubsection{Species mean-field approximation}
Here, we adopt the species mean-field (SMF) approximation, which assumes the wave function of the Bose-Fermi mixture to be of product form, i.e., $|\Psi \rangle = |\psi^{f}_{\text{SMF}}\rangle |\psi^{b}_{\text{SMF}}\rangle$, while excluding the interspecies correlations completely. Be aware that the SMF description still allows for arbitrarily large intraspecies correlations, which is of course beyond the mean-field approximation for mixtures, since the latter simply assigns a single permanent (Slater determinant) to bosonic (fermionic) species, while keeping the total wave function of the form as $|\Psi \rangle = |\psi^{f}_{\text{MF}}\rangle |\psi^{b}_{\text{MF}}\rangle$ \cite{BEC_Pethick,BEC_Stringari}. Variation of the species state $|\psi^{b}_{\text{SMF}} \rangle $ immediately allows one to derive the effective Hamiltonian for bosonic species as $\hat{H}^{b}_{\text{eff-SMF}} = \hat{H}_{b} + \hat{V}^{b}_{\text{SMF}} $. It contains, besides the original single-species Hamiltonian $\hat{H}_{b}$, an additional induced potential given by
\begin{align}
\hat{V}^{b}_{\text{SMF}} &= \langle \psi_{\text{SMF}}^{f}|\hat{H}_{bf} |\psi_{\text{SMF}}^{f}\rangle \nonumber \\
 &= \int dx ~\textit{V}^{b}_{\text{SMF}}(x) ~ \hat{\psi}^{\dagger} (x) \hat{\psi} (x),  \label {induced_potential_SMF}
\end{align}
with $\textit{V}^{b}_{\text{SMF}}(x) = g_{bf} \rho_{1-\text{SMF}}^{f} (x)$ and $ \rho_{1-\text{SMF}}^{f} (x) = \langle \psi_{\text{SMF}}^{f}| \hat{\psi}^{\dagger}_{f} (x)\hat{\psi}_{f} (x) |\psi_{\text{SMF}}^{f}\rangle$ being the SMF induced potential and the reduced one-body density for fermonic species obtained from the SMF simulations. At this point, we conclude that the SMF description incorporates the impact of the interspecies interactions onto the bosonic species as a mean-field induced potential. Albeit the fact that the SMF description is only exact for $\kappa = 0$, it can be qualitatively valid in the case when $\kappa$ is small enough. Compared to the \textit{ab~initio} ML-MCTDHX results, the $P$ value obtained from the SMF approximation possesses a qualitative agreement in the regimes of $P < 0$ (results are not shown here). However, large discrepancies occur for small $g_{b}$, in particular, when $g_{bf} \gg g_{b}$ [cf. Fig.~\ref{P_diff} (gray dotted lines)]. Note that, the presented $P$ values using the SMF description are for $g_{b} = 0.0,0.4,0.8$. For the other cases, both methods have minor discrepancies.

\begin{figure*}
  \centering
  \includegraphics[width=\textwidth]{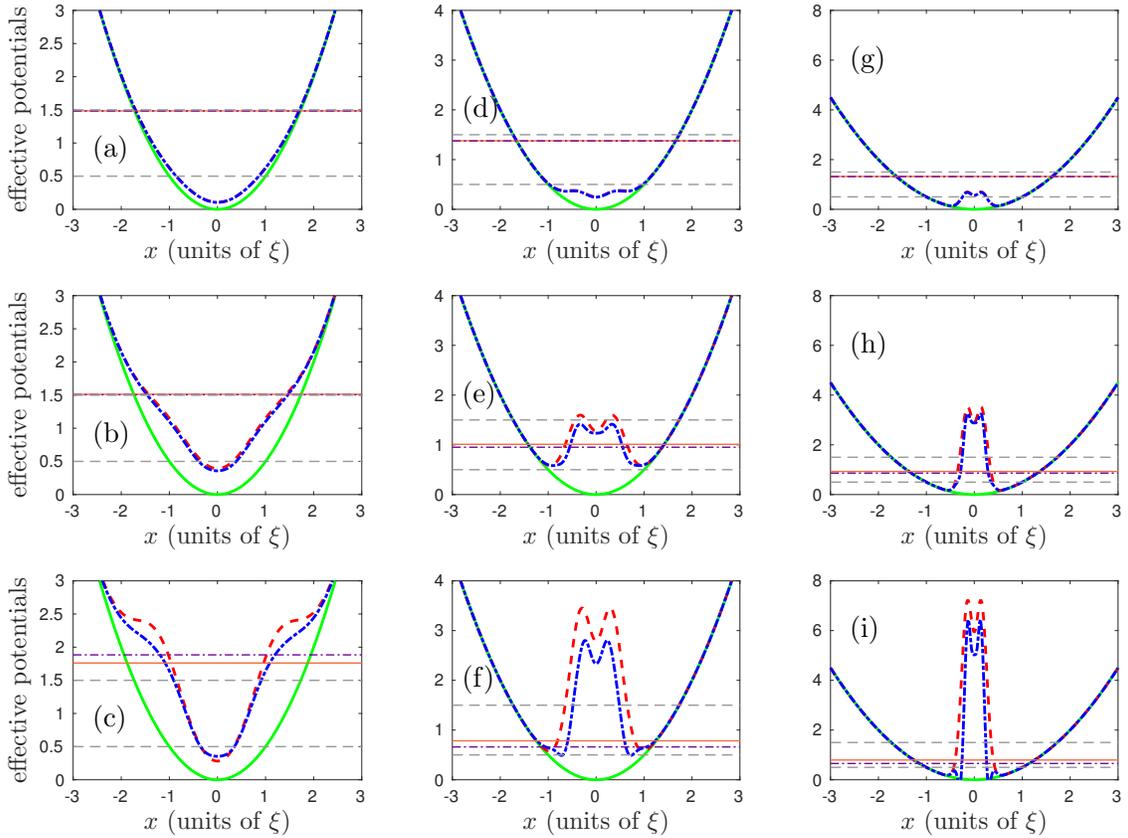}\hfill
  \caption{ Effective potentials (in units of $\eta$) for the bosonic species for $g_{b} = 0$ and $g_{bf} = 0.2$ (upper panels), $1.0$ (middle panels), and $2.0$ (lower panels). Besides, the left, middle, and right column belongs to the mass ratio $\beta = 1$, $\beta=5$, and $\beta=25$, respectively. The green solid lines represent the harmonic trap; the red dashed lines as well as the blue dash-dot lines are the profiles of $\textit{V}_{\text{eff-SMF}}^{b}(x)$ and $\textit{V}_{\text{eff}}^{b}(x)$. Furthermore, the straight horizontal lines indicate the lowest two single-particle energy levels [$\epsilon_{1,2}$ (in units of $\eta$)] for the harmonic trap (gray dashed lines), the SMF effective potential (purple dash-dot lines), and the $\textit{V}_{\text{eff}}^{b}(x)$ (brown solid lines). For illustrational purposes, we shift the energy levels of the $\textit{V}_{\text{eff-SMF}}^{b}(x)$ as well as of the $\textit{V}_{\text{eff}}^{b}(x)$ such that the ground-state energy matches the one of the harmonic trap ($\epsilon_{1} = 0.5$).}
\label{effective_potentials}
\end{figure*}

Since the SMF description reduces the interspecies physics to an additional potential, we investigate the SMF effective potential as $\textit{V}_{\text{eff-SMF}}^{b}(x) = \frac{1}{2}x^{2} + \textit{V}_{\text{SMF}}^{b}(x)$, which depicts the net confinement that a boson feels, and present it in Fig.~\ref{effective_potentials} (red dashed lines). For increasing $g_{bf}$, we observe that the SMF effective potential deviates significantly from the original harmonic confinement (green solid lines), forming either a tighter confinement for $\beta = 1$, [cf. Figs.~\ref{effective_potentials}(a)--\ref{effective_potentials}(c)] or a double-well pattern for $\beta = 5,25$ [cf. Figs.~\ref{effective_potentials}(d)--\ref{effective_potentials}(f) and \ref{effective_potentials}(g)--\ref{effective_potentials}(i)]. Here, we note that the changes of $g_{b}$ ($ g_{b} \in [0,2]$) only slightly affect the shapes of the SMF effective potential. In comparison to the harmonic trap, a tighter confinement enlarges the energy difference between the lowest two single-particle energy levels $\delta \epsilon= \epsilon_{2} - \epsilon_{1}$ \footnote{For illustrational purposes, we shift the energy levels of the $\textit{V}_{\text{eff-SMF}}^{b}(x)$ and the $\textit{V}_{\text{eff}}^{b}(x)$ such that the ground-state energy matches the one of the harmonic trap ($\epsilon_{1} = 0.5$). } [cf. Fig.~\ref{effective_potentials} (brown solid lines)] , which suppresses excitations of bosons by the intraspecies repulsion. Hence, for $\beta = 1$, the effective potential always reinforces the bosonic coherence leading to an increase of the $P$ value [cf. Fig.~\ref{P_diff}(a)]. In contrast, in the unequal mass cases, a double-well potential suppresses the particle hopping between two sides of the effective potential due to the presence of the central barrier. For increasing $g_{bf}$, the barrier height grows correspondingly, reducing $\delta \epsilon$ and facilitating the bosonic antibunching due to the bosonic repulsion. This can be clearly seen by the continuous decrease of the $P$ value in the mass-imbalanced regimes [cf. Figs.~\ref{P_diff}(b) and \ref{P_diff}(c)]. 

It is worth noting that, since the SMF description only results in an effective potential for the bosons, the repulsive intraspecies interaction restricts the two-body density imbalance to $P < 0$, which can be clearly seen in Fig.~\ref{P_diff} (gray dotted lines). Consequently, the SMF description fails to account for the occurrence of the bosonic bunching.

\subsubsection{Beyond the SMF description}
In order to go beyond the SMF description and understand the role of the interspecies entanglement on the bunching-antibunching crossover, we adopt the approach developed in Ref. \cite{effective_theory}, which incorporates the interspecies correlations to first order into the effective single-species description. The resulting effective Hamiltonian for $\sigma$ species is
\begin{equation}
\hat{H}_{\text{eff}}^{\sigma} = \textit{H}_{11}^{\bar{\sigma}} + \sum_{i \neq 1} \frac{ \sqrt{\lambda_{i}} \textit{H}_{1i}^{\bar{\sigma}} \textit{H}_{i1}^{\bar{\sigma}}}{\textit{t}_{1i}}, \label{effctive_Hamiltonian}
\end{equation}
with $\bar{\sigma} = f(b)$ for $\sigma = b(f)$. Moreover, $\textit{H}_{1i}^{\bar{\sigma}} = \langle \psi_{1}^{\bar{\sigma}}|\hat{H}|\psi_{i}^{\bar{\sigma}}\rangle$ and $\textit{t}_{1i} = \langle \psi_{1}^{\sigma}| \langle \psi_{1}^{\bar{\sigma}}|\hat{H}|\psi_{i}^{\bar{\sigma}}\rangle |\psi_{i}^{\sigma} \rangle $ representing the transition amplitude between the Schmidt-state products $ |\psi_{1}^{\bar{\sigma}}\rangle |\psi_{1}^{\sigma} \rangle$ and $ |\psi_{i}^{\bar{\sigma}}\rangle |\psi_{i}^{\sigma} \rangle$. It is worth noting that such an effective description focuses on the weak-entanglement regime described by the conditions $\sqrt{\lambda_{1}} \approx 1$ and $\sqrt{\lambda_{i \neq 1}} \ll 1$. In this spirit, the first Schmidt state contains the dominant contribution to the properties of the many-body state while all the terms of order $(\sqrt{\lambda_{i \neq 1}})^{2}$ are negligible. In contrast to the SMF effective Hamiltonian, $\hat{H}_{\text{eff}}^{\sigma} $ contains an additional interaction effectively present among the particles of the same type which originates from the interspecies correlations. For our Bose-Fermi mixture, we highlight that in particular for the case $g_{b}=0$, this induced interaction plays a crucial role (see below). We can rewrite the effective Hamiltonian for the bosonic species as
\begin{equation}
\hat{H}_{\text{eff}}^{b}  = \hat{H}_{b} + \hat{V}_{\text{ind}}^{b} + \hat{H}_{\text{ind}}^{b}, \label{eom_bf_1st_order}
\end{equation}
with
\begin{align}
\hat{V}_{\text{ind}}^{b} & = \int dx \left[ \textit{V}_{1}^{b}(x) + \textit{V}_{\text{no}}^{b}(x) \right]  \hat{\psi}^{\dagger}_{b}(x) \hat{\psi}_{b}(x) \label{v_eff_b},  \\
\hat{H}_{\text{ind}}^{b} &=  \frac{1}{2}\int dx_{1} dx_{2}~ \textit{H}_{\text{ind}}^{b} (x_{1}, x_{2}) \hat{\psi}^{\dagger}_{b}(x_{1}) \hat{\psi}^{\dagger}_{b}(x_{2}) \hat{\psi}_{b}(x_{2}) \hat{\psi}_{b}(x_{1}),\label{induced_interaction_b}
\end{align}
representing the induced potential and induced interaction, respectively. Here $\textit{V}_{1}^{b}(x)$ and $\textit{V}_{\text{no}}^{b}(x)$ stand for the contributions to the induced potential from the SMF approximation and the normal ordering of the induced interaction (see below). Note that we obtain the induced interactions and the induced potentials from the \textit{ab~initio} ML-MXTDHX simulations \cite{effective_theory, MX}. 
The computed two-body density imbalance for $g_{b} = 0, 0.4$ and $g_{b} = 0.8$ using the effective single-species Hamiltonian \eqref{eom_bf_1st_order} is presented in Fig.~\ref{P_diff} as well (colored dotted lines). Compared to the results obtained from the SMF approximation [cf. Fig.~\ref{P_diff} (gray dotted line)], the $P$ values using the beyond SMF description possess only minor discrepancy with the ones obtained from the \textit{ab~initio} ML-MCTDHX simulations. Importantly, the beyond SMF description successfully accounts for both the bunching and antibunching regimes; in particular, it quantitatively captures the critical bosonic repulsion $g_{b}^{c}$ in the mass-imbalanced regime, which manifests its applicability. We note again that the computed $P$ values from the SMF approximation, the beyond SMF description, and the \textit{ab~initio} ML-MCTDHX simulations have a good agreement for the other values of $g_{b}$ (results are not shown here).

Let us now inspect in more detail the induced potential and induced interaction in Eq.~\eqref{eom_bf_1st_order}. We first focus on the induced potential $\hat{V}_{\text{ind}}^{b}$, which consists of two terms where the first one 
\begin{equation}
\textit{V}_{1}^{b} (x)= g_{bf}  \gamma_{11}^{f} (x) \label{v_smf}
\end{equation}
represents the SMF contribution. Here $\gamma_{11}^{f}(x) = \langle\psi_{1}^{f}|\hat{\psi}^{\dagger}_{f} (x) \hat{\psi}_{f}(x) |\psi_{1}^{f}\rangle $ is the one-body transition matrix element for the fermionic species, which is the contribution from the first Schmidt state to the one-body density. Indeed, in the weak-entanglement regime we have $\gamma_{11}^{f}(x) \approx \rho_{1-\text{SMF}}^{f}(x) $; therefore, $\textit{V}_{1}^{b}(x)$ highly resembles the SMF induced potential (results are not shown here). The second term $\textit{V}_{\text{no}}^{b}(x)$ is given by
\begin{equation}
\textit{V}_{\text{no}}^{b} (x) = g_{bf} \sum_{i \neq 1} \frac{ \sqrt{\lambda_{i}}}{\tilde{t}_{1i}} \left[ \gamma_{1i}^{f}(x) \gamma_{i1}^{f} (x) + 2 \beta_{1i}^{f} \gamma_{i1}^{f}(x) \right], \label{v_ind}
\end{equation}
with $\tilde{t}_{1i}  =  \int dx~ \gamma_{1i}^{b}(x) \gamma_{1i}^{f}(x) $ and $\beta_{1i}^{f} = \langle\psi_{1}^{f}|\hat{H}_{f} |\psi_{i}^{f}\rangle$. Here the first part $\gamma_{1i}^{f}(x) \gamma_{i1}^{f}(x)$ is a result of the normal ordering of the term $\langle \psi_{1}^{f}|\hat{H}_{bf}|\psi_{i}^{f}\rangle \langle \psi_{i}^{f}|\hat{H}_{bf} |\psi_{1}^{f}\rangle $ in Eq.~\eqref{effctive_Hamiltonian}, while $\beta_{1i}^{f} \gamma_{i1}^{f}(x) $ stems from cross terms such as $\langle \psi_{1}^{f}|\hat{H}_{f}|\psi_{i}^{f}\rangle \langle \psi_{i}^{f}|\hat{H}_{bf}|\psi_{1}^{f}\rangle$. Note, however, that $\textit{V}_{\text{no}}^{b}(x)$ is a small correction compared to $ \textit{V}_{1}^{b}(x) $ due to the proportionality to $\sqrt{\lambda_{i \neq 1}}$. Similar to the SMF description, we introduce the effective potential as $\textit{V}_{\text{eff}}^{b}(x) = \frac{1}{2}x^{2} + \textit{V}_{\text{ind}}^{b}(x) $. We find that its profile highly resembles the one obtained from the SMF simulations incorporating in general only minor corrections [cf. Fig.~\ref{effective_potentials} (blue dash-dot lines)]. Interestingly, this correction always mitigates the effects of the induced potentials which are, hence, overestimated by the SMF description. This results in a narrower (broader) energy difference $\delta \epsilon$ in the mass-balanced (mass-imbalanced) regime [cf. Fig.~\ref{effective_potentials} (purple dash-dot lines and brown solid lines)].

\begin{figure*}
  \centering
  \includegraphics[width=\textwidth]{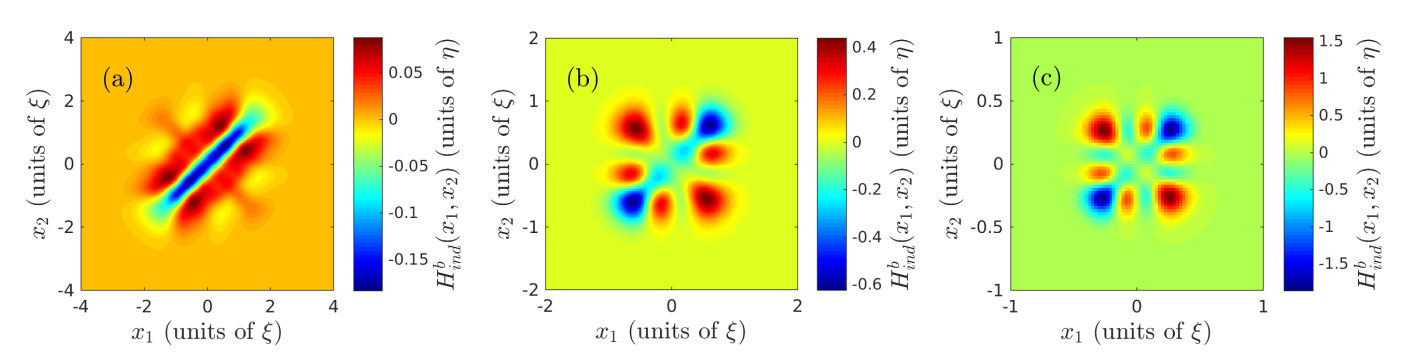}\hfill
  \caption{Induced interactions for the bosonic species for $g_{b} = 0$ and $g_{bf} = 1.0$ with (a) $\beta = 1$, (b) $\beta = 5$, and (c) $\beta = 25$, respectively.}
\label{induced_interactions}
\end{figure*}

Now we turn to the induced interaction, which reads 
\begin{equation}
\textit{H}_{\text{ind}}^{b}(x_{1}, x_{2}) = g_{bf}  \sum_{i \neq 1} \frac{ 2\sqrt{\lambda_{i}}}{\tilde{t}_{1i}} \gamma_{1i}^{f}(x_{1}) \gamma_{i1}^{f}(x_{2}). \label {ind-int_b}
\end{equation}
In Fig.~\ref{induced_interactions}, we present the induced interaction among the bosons for both the mass-balanced case and the mass-imbalanced cases. Importantly, the computed induced interaction preserves the particle exchange symmetry $ \textit{H}_{\text{ind}}^{b}(x_{1}, x_{2}) = \textit{H}_{\text{ind}}^{b}(x_{2}, x_{1})$ for indistinguishable particles as well as the parity symmetry $ \textit{H}_{\text{ind}}^{b}(x_{1}, x_{2}) = \textit{H}_{\text{ind}}^{b}( -x_{1}, -x_{2}) $ stemming from the original Hamiltonian [see Eq.~\eqref{Hamiltonian_bf}]. Moreover, we observe that, unlike the original zero-range bosonic repulsion, the induced interaction is long ranged and becomes, depending on the relative coordinate $r = x_{1} - x_{2}$, attractive for small particle distances $r$, repulsive for larger $r$, and vanishes at large $r$. Besides, the induced interaction also depends on the center-of-mass coordinate $R = (x_{1} + x_{2})/2$ owing to the inhomogeneity of the system. These features are drastically different from the situation in homogenous systems, where only the relative coordinate is involved and the induced interaction is overall attractive \cite{ind_cold_bf, ind_cold_bb}. Furthermore, we stress that, albeit the similar features, the induced interactions for various mass ratios differ from each other with respect to their strength and range due to the localization of the fermionic density. For instance, as compared to the equal mass case, the induced interaction for $\beta=25$ has an almost twice as large maximal value while being of smaller range.

The induced interaction paves the way to qualitatively exploit the physical impact of the interspecies correlations ignored by the SMF description. In particular, it enables us to understand the occurrence of the bosonic bunching. Based on its profile, we realize that the attractive part of the induced interaction supports the configuration that one boson stays in the vicinity of the other and even suppresses configurations with both bosons being apart from each other. Based on the above two aspects, we conclude that the induced interaction enforces the bunching for the bosonic species. These findings result in the following important outcomes. (i) For the case with $g_{b} = 0$, the induced interaction among the bosons leads to the rising trend of the two-body density imbalance. (ii) For the case of a nonvanishing $g_{b}$, there exists a competition between the induced interaction and the contact bosonic repulsion. For small enough $g_{b}$, the induced interaction is dominant thereby resulting in a similar behavior of $P$ as the case for $g_{b} = 0$. However, once $g_{b}$ exceeds a critical value, the net interaction between two bosons becomes repulsive, leading to a continuous decrease of the two-body density imbalance in the mass-imbalanced cases. Note that this competition is not captured in the mass-balanced regime for the parameter regime under investigation. (iii) Moreover, the strength and range of the induced interaction highly depends on the mass ratio, therefore, resulting in a strong mass imbalance dependency of the critical bosonic repulsion $g_{b}^{c}$.

\subsection{Two-site extended Bose-Hubbard model} 
The above effective descriptions introduce significant insights in the study of mixtures while, at the same time, enable us to explore the physics effectively present in a single species. This is why we have been able to see that the effective potential in the strong mass-imbalanced regimes becomes a double well [cf. Figs.~\ref{effective_potentials}(g)--\ref{effective_potentials}(i)], which offers the opportunity to map the effective Hamiltonian for the bosonic species to a lattice model. In order to elucidate this lattice model, we focus on the situation for $\beta = 25$ in the following discussions. Furthermore, in Figs.~\ref{effective_potentials}(g)--\ref{effective_potentials}(i), we clearly see that the presence of a central barrier induces the formation of bands with a small energy spacing $\delta\epsilon$ between the two lowest single-particle eigenstates. Importantly, a large gap to the next band severely suppresses particle excitations to the higher bands. In this way, the bosons mainly populate the lowest band. With this knowledge, we adopt the single-band approximation and obtain the Hamiltonian for the two-site extended Bose-Hubbard (EBH) model for the two bosons as (for a corresponding derivation please see the Appendix)
\begin{align}
\hat{H}_{\text{EBH}} = &-J (\hat{a}^{\dagger}_{L}\hat{a}_{R} + \hat{a}^{\dagger}_{R} \hat{a}_{L} ) + V \hat{N}_{L} \hat{N}_{R} \nonumber \\
&+ \frac{U}{2} \left[\hat{N}_{L}(\hat{N}_{L}-1) + \hat{N}_{R}(\hat{N}_{R}-1) \right] \label{EBH_model}
\end{align}
where $\hat{a}^{\dagger}_{R/L}$ ($ \hat{a}_{R/L}$) denote the creation (annihilation) operators for the right or left site and the coefficients $J$, $U$, and $V$ represent the hopping amplitude and the on-site interaction, as well as the nearest-neighbor interaction, respectively. Compared to the conventional Bose-Hubbard model, the EBH model possesses richer phases including the density-wave phases and the Haldane insulator phase, in addition to the superfluid and the Mott insulating phase \cite{EBH_nearest_1,EBH_nearest_2}. 

Before proceeding, it is instructive to note that the discussions on the two-site EBH model as well as the computed coefficients are in the framework of the beyond SMF description. Let us first briefly comment on the roles of both $g_{b}$ and $g_{bf}$ with respect to the above two-site EBH model. The role of $g_{b}$ is relatively simple since the increment of the bosonic repulsion mainly increases the on-site repulsion, resulting in the increase of the $U$ value. In comparison, the role of $g_{bf}$ is more complicated. On one hand, the increase of $g_{bf}$ leads to a rapid increase of the height of the central barrier which, in turn, leads to a monotonous decrease of the hopping amplitude $J$. On the other hand, it also increases the strength of the induced interaction resulting in an increase of the on-site attraction, as well as the off-site repulsion. 

\begin{figure}
  \centering
  \includegraphics[width=0.5\textwidth]{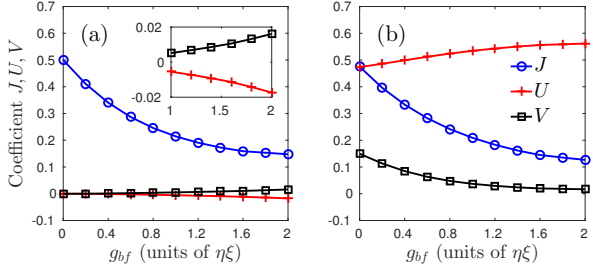}\hfill
  \caption{ Coefficients (in units of $\eta$) for the two-site EBH model for $\beta=25$. (a) $g_{b} = 0$; (b) $g_{b} = 1.0$. The inset shows the on-site and nearest-neighbor interaction for $g_{b} = 0$ and $g_{bf} \in [1,2]$. Results are obtained from the \textit{ab~initio} ML-MCTDHX simulations.}
\label{parameters for BH model}
\end{figure}

The computed coefficients for the two-site EBH model are presented in Fig.~\ref{parameters for BH model}. As anticipated, the increase of $g_{bf}$ leads to a decrease of the hopping amplitude $J$ since the increment of the lattice depth severely suppresses the particle hopping between the two sites. In contrast, the coefficients of both on-site and nearest-neighbor interactions are defined by the interplay between the induced interaction and the bosonic contact repulsion. For the case $g_{b} = 0$ [cf. Fig.~\ref{parameters for BH model}(a)], the induced interaction creates attractive on-site and repulsive nearest-neighbor interactions due to the long-range behavior of the spatial profile [cf. Fig.~\ref{induced_interactions}(c)]. Importantly, the on-site attraction combined with a weak hopping amplitude facilitates the ``cat-state-like" phase \cite{cat_state}, with the wave function $|\Psi \rangle  = 1/\sqrt{2}~ (|2,0\rangle +|0,2\rangle)$, which corresponds to the bosonic bunching. Here $|N_{L},N_{R} \rangle = 1/\sqrt{N_{L}!\,N_{R}!} (\hat{a}_{L}^{\dagger})^{N_{L}}(\hat{a}_{R}^{\dagger})^{N_{R}}|vac\rangle$, with $N_{L(R)}$ being the particle number of the left (right) site under the constraint of particle conservation $N_{L}+N_{R} = 2$. In contrast, for the cases with a large bosonic repulsion [cf. Fig.~\ref{parameters for BH model}(b)], the on-site repulsion dominates and, in turn, supports the formation of a ``Mott-state-like" phase, which corresponds to the antibunching of the bosons. As a result, the above two-site EBH model supports a transition from the cat-state-like phase to the Mott-state-like phase, which coincides with the bunching-antibunching crossover of the more general binary mixture system investigated above.

\begin{figure}
  \centering
  \includegraphics[width=0.5\textwidth]{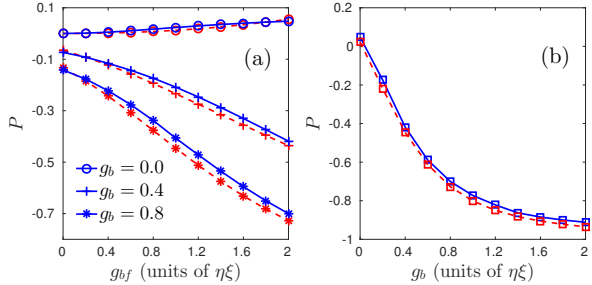}\hfill
  \caption{ Computed two-body density imbalance for $\beta = 25$ using the two-site EBH model (red dashed lines) and the \textit{ab~initio} ML-MCTDHX simulations (blue solid lines) for (a) $ g_{b} = 0.0,0.4,0.8$ and $g_{bf} \in [0,2]$ and (b) $g_{bf} = 2.0$ and $g_{b} \in [0,2]$.}
\label{P_MX_EBH}
\end{figure}

In order to quantitatively judge the validity of the two-site EBH model, we inspect the corresponding two-body density imbalance 
\begin{align}
P_{\text{EBH}} = &\sum_{i = L,R} \frac{1}{N_{L}!\,N_{R}!} \langle \Psi_{\text{EBH}} |\hat{a}_{i}^{\dagger} \hat{a}_{i}^{\dagger} \hat{a}_{i} \hat{a}_{i} |\Psi_{\text{EBH}} \rangle \nonumber \\
&-\langle \Psi_{\text{EBH}} |\hat{a}_{L}^{\dagger} \hat{a}_{R}^{\dagger} \hat{a}_{R} \hat{a}_{L} |\Psi_{\text{EBH}} \rangle,
\end{align}
denoting the probability difference between finding two bosons on the same site and finding them on different sites. Moreover, the ground state $|\Psi_{\text{EBH}} \rangle$ is obtained by diagonalizing the Hamiltonian $\hat{H}_{\text{EBH}}$ in the space spanned by the basis $\{|2,0\rangle, |1,1\rangle, |0,2\rangle \}$. In Fig.~\ref{P_MX_EBH}(a), we present the computed $P_{\text{EBH}}$ for $g_{b} = 0.0,0.4,0.8$ (red dashed lines). As compared to the results obtained from the \textit{ab~initio} ML-MCTDHX simulations (blue solid lines), the $P_{\text{EBH}}$ values possess only minor discrepancies. Importantly, using the two-site EBH model one can even quantitatively capture the critical bosonic repulsion $g_{b}^{c}$ that we discussed before, which manifests its applicability. Meanwhile, we also clearly see that the obtained lattice model successfully accounts for the transition from bunching to antibunching with increasing bosonic repulsion [cf. Fig.~\ref{P_MX_EBH}(b)]. 

\section{Conclusions and Outlook} \label{Conclusions}
We have investigated a few-body mixture of ultracold bosons and fermions in a one-dimensional harmonic confinement. In particular, we focus on the analysis and discussion of the ground-state properties for various interaction strengths and mass ratios. By employing the \textit{ab~initio} ML-MCTDHX approach, we obtain the ground-state wave function including all correlations. We first present our main observation of the bunching-antibunching crossover of the bosonic species induced by the interspecies correlations, which can be observed via the bosonic two-body density. Particularly, in the mass-imbalanced regimes, we observe the existence of a critical value of the bosonic repulsion, below or above which, the increment of Bose-Fermi repulsion will lead the bosons into a bunching or antibunching regime.

In order to unveil the physical origin of this crossover, we apply two approximate methods. First, we adopt a species mean-field description which excludes (includes) all the interspecies (intraspecies) correlations and incorporates the impact of the fermionic species into a mean-field induced potential. Albeit the SMF description can qualitatively account for the antibunching regime, it fails to describe the bunching regime. Second, we employ a beyond SMF description, which accounts for the interspecies correlations to first order and results in an effective single-species Hamiltonian containing, besides the induced potential, an additional induced bosonic interaction. Such an induced interaction enables us to understand the emergence of the bosonic bunching. Finally, in the strongly mass imbalanced regime, we derive a two-site extended Bose-Hubbard model which accounts for the low-energy effective physics of the bosons.

Our work presents the rich physics of the 1D few-body Bose-Fermi mixture. In particular, the provided effective single-species descriptions allow for gaining physical insights into the mechanisms behind certain observations which offers an intriguing approach for the studies of mixtures. Concerning future investigations, it is of specific interest how the system properties, in particular the bunching-antibunching crossover, depend on the spatial dimensions and in particular the particle number. Moreover, it is also interesting to analyze the form of the induced interaction depending on the particle statistics and the underlying interspecies coupling. 

\begin{acknowledgments}
The authors acknowledge fruitful discussions with Sven Kr\"onke, Kevin Keiler, and Maxim Pyzh. J.C. and P.S. gratefully acknowledge financial support by the Deutsche Forschungsgemeinschaft (DFG) in the framework of the SFB 925 ``Light induced dynamics and control of correlated quantum systems". The excellence cluster ``The Hamburg Centre for Ultrafast Imaging-Structure: Dynamics and Control of Matter at the Atomic Scale" is acknowledged for financial support.
\end{acknowledgments}

\section{Appendix I: Derivation of the two-site extended Bose-Hubbard model}
We present here the detailed derivations for the two-site EBH model provided in Eq.~\eqref{EBH_model}. According to Eq.~\eqref{eom_bf_1st_order}, the effective Hamiltonian for the bosonic species is given by
\begin{align}
\hat{H}_{\text{eff}}^{b} &=\int dx~\hat{\psi}^{\dagger}_{b}(x) h_{\text{eff}}^{b}(x) \hat{\psi}_{b}(x) \nonumber \\
&+\frac{g_{b}}{2}\int dx~\hat{\psi}^{\dagger}_{b}(x)\hat{\psi}^{\dagger}_{b}(x)\hat{\psi}_{b}(x)\hat{\psi}_{b}(x) \nonumber\\
&+\frac{1}{2}\int dx_{1} dx_{2}~ \textit{H}_{\text{ind}}^{b} (x_{1}, x_{2}) \hat{\psi}^{\dagger}_{\sigma}(x_{1}) \hat{\psi}^{\dagger}_{\sigma}(x_{2}) \hat{\psi}_{\sigma}(x_{2}) \hat{\psi}_{\sigma}(x_{1}) \label {eom_eff_b_appendix}
\end{align}
here $h_{\text{eff}}^{b}(x) = -\frac{1}{2}\frac{\partial^{2}}{\partial x^{2}} + V_{\text{eff}}^{b}(x)$ is the single-particle Hamiltonian with the effective potential. To obtain the two-site extended Bose-Hubbard model, we first expand the field operator as
\begin{equation}
\hat{\psi}_{b}(x) = \phi_{1}(x) \hat{a}_{1} + \phi_{2}(x) \hat{a}_{2}, \label{2-mode expand psi}
\end{equation}
with $\phi_{1}(x)$ [$\phi_{2}(x)$] being the single-particle ground (first excited) state of the effective potential $V_{\text{eff}}^{b}(x)$. Substituting Eq.~\eqref{2-mode expand psi} into Eq.~\eqref{eom_eff_b_appendix} yields  
\begin{align}
\hat{H}_{\text{eff}}^{b} &= \sum_{i = 1}^{2} ~\varepsilon_{i} \hat{N}_{i} + P_{i} (\hat{N}_{i}^{2} - \hat{N}_{i}) \nonumber \\
& + P_{12} \left[(\hat{a}_{1}^{\dagger} \hat{a}_{2})^{2}  + (\hat{a}_{2}^{\dagger} \hat{a}_{1})^{2}  \right] + P_{3} \hat{N}_{1}\hat{N}_{2} \label{EBH_model_bloch}
\end{align}
with
\begin{align}
\varepsilon_{i} &= \int dx~ \phi_{i}(x) h_{\text{eff}}^{b}(x)  \phi_{i}(x) \nonumber \\
T_{i} &= \frac{g_{b}}{2} \int dx~ \phi_{i}^{4}(x) + \textit{H}_{iiii}^{b} \nonumber \\
T_{12} &= \frac{g_{b}}{2} \int dx~ \phi_{1}^{2}(x) \phi_{2}^{2}(x) + \textit{H}_{1122}^{b} \nonumber \\
T_{3} &= 2 \left[ g_{b}\int dx~ \phi_{1}^{2}(x) \phi_{2}^{2}(x) + \textit{H}_{1212}^{b} + \textit{H}_{1221}^{b} \right] \nonumber \\
\end{align}
and
\begin{equation}
\textit{H}^{b}_{ijkl} = \frac{1}{2}\int dx_{1} dx_{2}~ \phi_{i}(x_{1}) \phi_{j}(x_{2}) \textit{H}_{\text{ind}}^{b} (x_{1}, x_{2}) \phi_{k}(x_{1}) \phi_{l}(x_{2}).
\end{equation}
We emphasize that the terms such as $\textit{H}^{b}_{iiij}$ ($i \neq j$) are eliminated by making use of the parity symmetry of the induced interaction.

Next, we project the Hamiltonian \eqref{EBH_model_bloch} onto the Wannier basis, which is the linear combinations of $\phi_{1}$ and $\phi_{2}$ as $\phi_{L/R}(x) = \frac{1}{\sqrt{2}} \left[\phi_{1}(x) \pm \phi_{2}(x) \right]$, and finally we arrive at
\begin{align}
\hat{H}_{\text{EBH}} = &-J (\hat{a}^{\dagger}_{L}\hat{a}_{R} + \hat{a}^{\dagger}_{R} \hat{a}_{L} ) - \Omega \left[ (\hat{a}_{L}^{\dagger} \hat{a}_{R})^{2} + (\hat{a}_{R}^{\dagger} \hat{a}_{L})^{2} \right] \nonumber \\
&+ \frac{U}{2} \left[\hat{N}_{L}(\hat{N}_{L}-1) + \hat{N}_{R}(\hat{N}_{R}-1) \right] + V \hat{N}_{L} \hat{N}_{R}
\end{align}
with
\begin{align}
J &= \frac{\varepsilon_{2}-\varepsilon_{1}}{2} + \frac{T_{2} - T_{1}}{2} (N_{b}-1) \nonumber \\
U &= \frac{T_{1}+T_{2}+2T_{12}+T_{3}}{2} \nonumber \\
\Omega & = \frac{T_{1}+T_{2}+2T_{12}-T_{3}}{4} \nonumber \\
V & = T_{1} + T_{2} -2T_{12}
\end{align}
being the coefficients for the hopping amplitude, on-site interaction, and pair-tunneling amplitude, as well as the nearest-neighbor interaction, respectively. It is worth noting that the amplitude of the pair tunneling $\Omega$ is one order of magnitude smaller as compared to the other coefficients, and we therefore omit it in Eq.~\eqref{EBH_model}.

\end{document}